# Multimodal AI-driven Biomarker for Early Detection of Cancer Cachexia

**Sabeen Ahmed** [1,7, *], **Nathan Parker** [2], **Margaret Park** [3,6], **Evan W. Davis** [3,5], **Jennifer B. Permuth** [3,5], **Matthew B. Schabath** [5], **Yasin Yilmaz** [7], **and Ghulam Rasool** [1,7]

[1] Department of Machine Learning, H. Lee Moffitt Cancer Center and Research Institute, Tampa, FL; sabeen.ahmed@moffitt.org, ghulam.rasool@moffitt.org

[2] Department of Health Outcomes and Behavior, H. Lee Moffitt Cancer Center and Research Institute, Tampa, FL; nathan.parker@moffitt.org

[3] Department of GI Oncology, H. Lee Moffitt Cancer Center and Research Institute, Tampa, FL; margaret.park@moffitt.org, evan.davis@moffitt.org, jenny.permuth@moffitt.org

[4] Diagnostic Imaging and Interventional Radiology, H. Lee Moffitt Cancer Center and Research Institute, Tampa, FL; daniel.jeong@moffitt.org

[5] Department of Cancer Epidemiology, H. Lee Moffitt Cancer Center and Research Institute, Tampa, FL; lauren.peres@moffitt.org, evan.davis@moffitt.org, jenny.permuth@moffitt.org, matthew.schabath@moffitt.org

[6] Department of Biostatistics and Bioinformatics, H. Lee Moffitt Cancer Center and Research Institute, Tampa, FL; margaret.park@moffitt.org

[7] Department of Electrical Engineering, University of South Florida, Tampa, FL; yasiny@usf.edu, sabeen.ahmed@moffitt.org, ghulam.rasool@moffitt.org

[8] Epidemiology and Genomics Research Program, National Cancer Institute, NIH; erin.siegel@nih.gov

* Corresponding author: sabeen.ahmed@moffitt.org

**Abstract:** Cancer cachexia is a multifactorial syndrome characterized by progressive muscle wasting, metabolic dysfunction, and systemic inflammation, leading to reduced quality of life and increased mortality. Despite extensive research, no single definitive biomarker exists, as cachexia-related indicators such as serum biomarkers, skeletal muscle measurements, and metabolic abnormalities often overlap with other conditions. Existing composite indices, including the Cancer Cachexia Index (CXI), Modified CXI (mCXI), and Cachexia Score (CASCO), integrate multiple biomarkers but lack standardized thresholds, limiting their clinical utility. This study proposes a multimodal AI-based biomarker for early cancer cachexia detection, leveraging open-source large language models (LLMs) and foundation models trained on medical data. The approach integrates heterogeneous patient data, including demographics, disease status, lab reports, radiological imaging (CT scans), and clinical notes, using a machine learning framework that can handle missing data. Unlike previous AI-based models trained on curated datasets, this method utilizes routinely collected clinical data, enhancing real-world applicability. Additionally, the model incorporates confidence estimation, allowing the identification of cases requiring expert review for precise clinical interpretation. Preliminary findings demonstrate that integrating multiple data modalities improves cachexia prediction accuracy at the time of cancer diagnosis. The AI-based biomarker dynamically adapts to patient-specific factors such as age, race, ethnicity, weight, cancer type, and stage, avoiding the limitations of fixed-threshold biomarkers. This multimodal AI biomarker provides a scalable and clinically viable solution for early cancer cachexia detection, facilitating personalized interventions and potentially improving treatment outcomes and patient survival.

1. **Introduction**

   Cancer cachexia is a multifactorial syndrome associated with poor quality of life and survival. The incidence of cachexia varies with the type of cancer, with gastroesophageal and pancreatic cancers having the highest rate of around 60-70%, followed by 40-50% for hematological, colorectal, and lung cancers [1].

   Cachexia results from a complex interplay of metabolic changes, chronic inflammation, hormonal imbalances, reduced food intake, and altered muscle metabolism. Several serum biomarkers are used to assess these underlying factors, including elevated C-reactive protein (CRP) as a marker of systemic inflammation, and low serum albumin as an indicator of malnutrition [2] [3]. However, these biomarkers are not exclusive to cachexia and can be influenced by other conditions, reducing their efficacy. Similarly, skeletal muscle assessment via radiological imaging, particularly through skeletal muscle index (SMI), is a characteristic and widely used metric, but muscle loss can also occur temporarily due to surgery and treatment effects. To overcome these challenges, composite indices such as the cachexia index (CXI) and modified cachexia index (mCXI) have been developed, integrating SMI, nutritional status, and inflammatory markers [4] [5]. These indices have shown potential as biomarkers for cancer cachexia, but they lack standardized thresholds for clinical use. Another comprehensive tool, the cachexia score (CASCO), evaluates multiple domains such as metabolic disturbances, inflammation, and physical performance, to provide a holistic assessment of the patient's condition [6]. However, CASCO is resource-intensive, requiring extensive measurements and patient questionnaires, making it impractical for routine clinical use and rapid screening. Despite these advancements, a universally accepted, independent biomarker for cancer cachexia remains elusive, underscoring the need for a more integrative and scalable approach.

   To address these limitations, we propose a multimodal AI-based biomarker for early detection of cancer cachexia, designed specifically for clinical application. Our approach integrates diverse patient data, including patient demographics, disease status, laboratory reports, imaging data, and clinical notes, and utilizes machine learning algorithms to detect cancer cachexia at the time of cancer diagnosis.  Unlike previous studies using AI for cachexia prediction, that relied on specially curated research cohorts [7], our approach leverages routinely collected clinical data, improving real-world applicability. Additionally, we incorporate model confidence estimation, allowing the model to flag cases requiring expert review, ensuring that clinical decision-making is both data-driven and reliable. Preliminary results indicate that integrating multiple data modalities enhances the accuracy of cachexia prediction at the time of diagnosis.

An advantage of our AI biomarker is that it is not limited to a fixed threshold. Instead, it dynamically learns and adapts thresholds based on patient-specific factors, including age, race, ethnicity, weight, cancer type, and stage. This adaptability ensures a personalized, data-driven approach to cachexia detection, making it more applicable across diverse patient populations. The development of a multimodal AI biomarker holds promise for quick, accurate, reliable, and early detection of cancer cachexia, potentially leading to timely interventions, personalized treatment strategies, and improved patient outcomes.

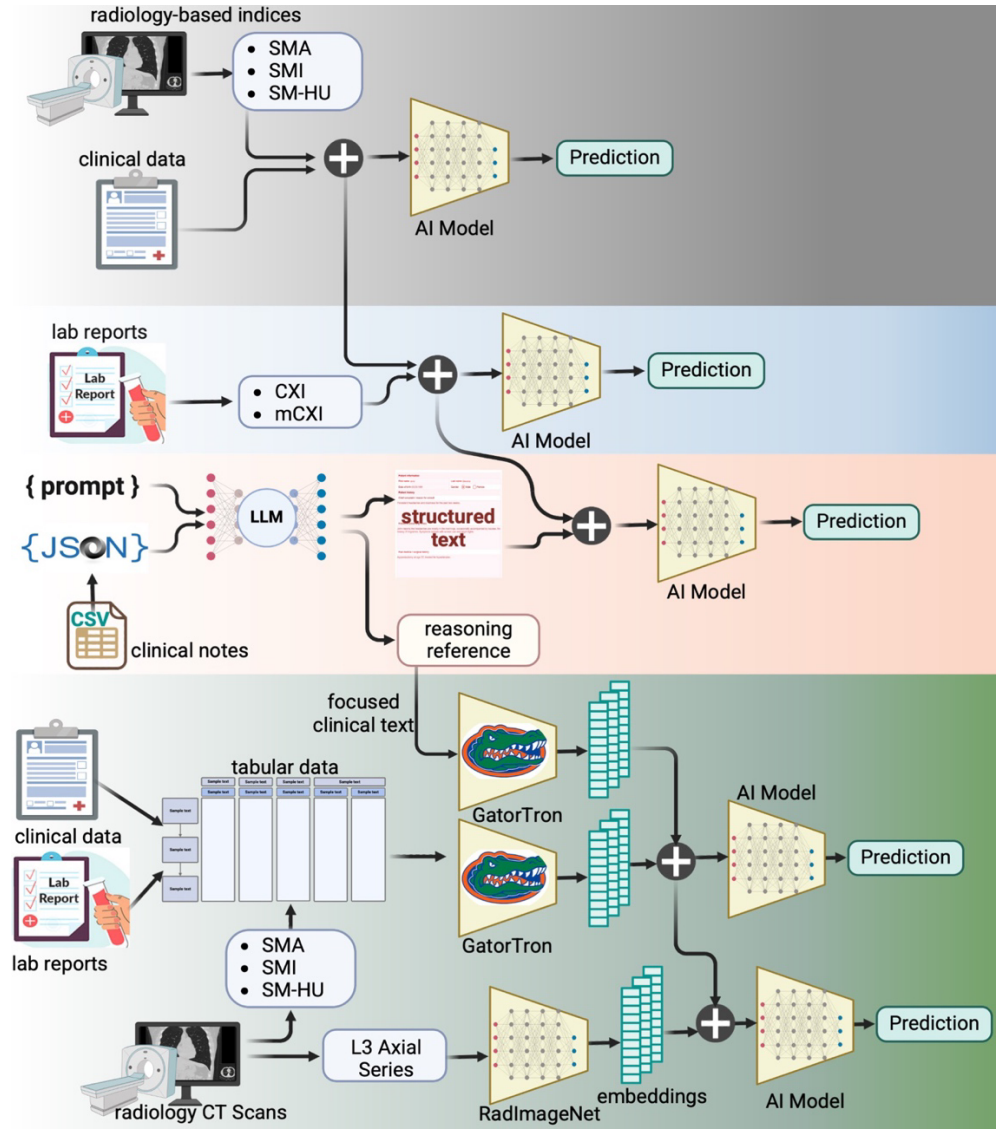

**Figure 1. Framework for Multimodal Data Integration in Cachexia Prediction.** Various combinations of data modalities are used to train the machine learning models to detect cachexia in cancer patients. These combinations include: **(a)** clinical data + CT images based skeletal muscle measurements (SM measurements); **(b)** data modalities in (a) with the addition of lab reports; **(c)** data modalities in (b) with the addition of structured clinical notes generated using Deepseek-70b LLM; **(d)** GatorTron to generate embeddings for clinical data + SM measurements + lab reports

in tabular format concatenated with embeddings for focused clinical notes; **(e)** embeddings generated in (d) concatenated with embeddings for CT image slices at L3 level from axial series generated by RadImageNet (a foundation model for radiology images). The confidence score is used to segregate incorrect and correct predictions with low model confidence for expert review. This framework can easily be integrated into clinical workflows.

## 2. Materials and Methods

*2.1. Datasets*

This study used a patient cohort from the Florida Pancreas Collaborative study [8] which has patients from various hospitals around Florida. We have built a cohort of PDAC patients only. Out of the 318 PDAC patients, 236 were selected based on data availability, including skeletal muscle area (SMA) and cachexia status. Of these 236 patients, 131 belonged to Moffitt Cancer Center (Figure 2). This study was approved by the Moffitt Cancer Center Scientific Review Committee and the Advarra Institutional Review Board.

This study included clinical data, radiology images, lab reports, and clinical notes at the time of cancer diagnosis. Demographics of the cohort and details of the clinical data used are given in Table 1 and Figure 2. Radiology data included CT scans from the time of diagnosis comprising DICOM images with all or a combination of axial, sagittal, coronal views, maximum intensity projections, contrast, non-contrast and within contrast, arterial, venous, and delayed phases. CT scans were only available for Moffitt patients. The SMA, SMI, and the average skeletal muscle Hounsfield unit (SM HU) were extracted from CT images using SMAART-AI for Moffitt patients and AW Server for outside Moffitt patients. The end of the third lumbar level slice from the venous phase series, when available, was used for estimating SMA, SMI, and average SM HU. The lab reports used in this study are given in Table 2. C-reactive protein (CRP) values were not included since these were not available for approximately 95% of the patients. Table 3 lists the clinical notes used in this study.

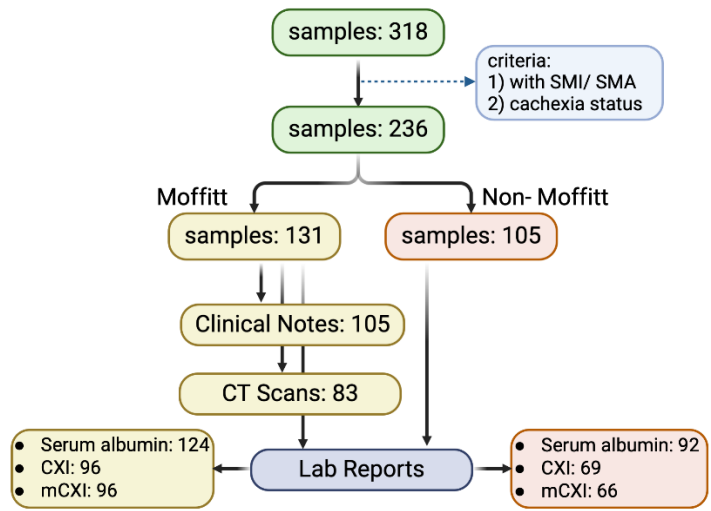

**Figure 2. PDAC patient's cohort branching into Moffitt and Non-Moffitt samples.** Samples with available clinical notes, CT scans, and lab reports are shown in each sub-branch of the Moffitt and non-Moffitt groups.

Table 1. PDAC patient's cohort and clinical data.

|  | Total patient count = 236 |
|---|---|
| Age at diagnosis, mean (SD) | 69.05 ± 10.13 |
| Weight (lbs.) at diagnosis, mean (SD) | 166.95 ± 37.88 |
| Height (m), mean (SD) | 1.69 ± 0.59 |
| BMI at diagnosis, mean (SD) | 26.75 ± 5.77 |
| Sex, N |  |
|   Female | 119 |
|   Male | 117 |
| Race and Ethnicity, N |  |
|   Non-Hispanic White | 176 |
|   Hispanic/Latinx | 36 |
|   Non-Hispanic Black | 24 |
| TNM Stage (Pathological), N |  |
|   1: IA (T1, N0, M0) | 15 |
|   2: IB (T2, N0, M0) | 27 |
|   3: IIA (T3, N0, M0 | 26 |
|   4: IIA (T1, N1, M0) | 1 |
|   5: IIA (T2, N1, M0) | 17 |
|   6: IIB (T3, N1, M0) | 16 |
|   7: III (T4, Any N, M0) | 29 |
|   8: IV (Any T, Any N, M1) | 71 |
|   9: Tumor stage cannot be assessed | 13 |
|   -1: NA | 21 |
| Cachexia Status, N |  |
|   Cachectic | 152 |
|   Non-cachectic | 84 |
| Patient Hospital, N |  |
|   Moffitt | 131 |
|   Outside Moffitt | 105 |

Table 2. Lab reports and derived metrics included in the study.

|  | Total patient count = 236 |
|---|---|
| Serum albumin (g/dL), N | 216 |
| Neutrophil count (absolute, k/uL), N | 167 |
| Lymphocyte count (absolute, k/uL), N | 170 |
| Serum Blood Urea Nitrogen (mg/dL), N | 221 |
| Serum Creatinine (mg/dL), N | 218 |
| Metrics used |  |
|   Neutrophil-to-Lymphocyte ratio (NLR), N | 167 |
|   Cachexia index (CXI), N | 165 |
|   Urea-to-Creatinine ratio (UCR), N | 217 |

**Table 3**. Types of clinical notes used in this study.

Nutrition assessment form
Nutrition diagnosis/comments
Progress note
Dietary assessment/evaluation/comments
Inter visit note
Ambulatory care note
History and physical note
Patient assessment

*2.2. Data Processing*

2.2.1. Clinical Data

The clinical data was processed with missing values for age imputed using the mean of the entire population. The missing values for weight and height were imputed using the mean of the population filtered based on the patient's sex and race/ethnicity. For cases where the tumor stage could not be assessed or was not available, '-1' was used to represent missing data. The qualitative values, such as gender, race/ethnicity, and cancer stage, were binarized.

Patients were divided into different cachexia categories using the following criteria defined in the literature:

1. The two-stage system (cachectic/non-cachectic) introduced by Fearon et al. [9] categorizes patients as either cachectic or non-cachectic. A patient was termed cachectic if either of the following conditions were met:
    a. Weight loss was >5% over the past six months for BMI ≥ 20.
    b. Weight loss was >2% over the past six months for a BMI < 20.
2. The four-stage system (pre-cachectic, cachectic, refractory, and non-cachectic) proposed by Vigano et al. [10]. This classification used five criteria:
    a. Biochemical markers (elevated C-reactive protein or leukocytes, hypoalbuminemia, or anemia)
    b. Food intake (normal or decreased)
    c. Moderate weight loss over the past six months (≤ 5%)
    d. Significant weight loss over the past six months (> 5%)
    e. Performance status (Eastern Cooperative Oncology Group Performance Status ≥ 3).

If none of these criteria were met, the patient was considered non-cachectic. When sufficient information was unavailable for the four-stage system, the two-stage system was used.

**Tabular data embeddings**: The clinical data in tabular format was converted into text for embeddings. The GatorTron-medium model, developed by the University of Florida in collaboration with NVIDIA [11], was utilized to generate embeddings from tabular clinical data. Missing values were replaced by the word 'missing' to add meaning. The original text was used for qualitative data.

2.2.2. Lab Reports

Lab reports listed in Table 2 were used to derive the cancer cachexia index (CXI) introduced by Jafri et al. [12], which has been shown to be a useful potential biomarker for cancer cachexia [4]. CXI is calculated using the following formula,

$$CXI = \frac{SMI \times serum\ albumin}{NLR}$$

where NLR is the ratio of absolute neutrophil count to absolute lymphocyte count. The modified cancer cachexia index (mCXI), introduced by Yuan et al. [5], was calculated using the following formula,

$$mCXI = \frac{serum\ albumin}{NLR \times UCR}$$

where UCR is the ratio of blood urea nitrogen to serum creatinine.

Lab values, along with intermediate metrics such as NLR and UCR, were used together with CXI and mCXI to represent lab report data. Lab values that were not available were replaced by '-1', indicating missing data. In case any component for calculating CXI or mCXI was missing, a '-1' was used to represent the unavailability of these indices.

2.2.3. Clinical Notes Processing

Clinical notes data was processed in two different ways,

1. Structured report format.
2. Embeddings for the textual data.

**Structured report format**: All available notes for each patient were combined in a JSON file. These textual data notes were processed using Llama3.2:3b-instruct-fp16, Qwen2.5:7b-instruct-fp16, and Deepseek-r1:70b LLMs running on local GPU machines served by Ollama. The LLM was used to extract responses in yes, no, and not-given format based on the notes available for each patient, for a set of questions relevant to determining cachexia status. The LLM was instructed to provide reasoning for its response along with references from the text. The response to these questions was converted into a tabular format, with '1' representing 'yes', '0' representing 'no', and '-1' representing 'not-given.'

**Embeddings for the textual data**: The GatorTron-medium model, was utilized to generate embeddings from the clinical notes data. To minimize noise in the embeddings, we used only the reasoning and reference text extracted by the LLM, rather than the full set of combined notes for each patient. For each question, the corresponding reasoning and reference text were concatenated per patient and processed using GatorTron.

Given the model's 512-token input context limit, the text for each patient was segmented into 512-token chunks before embedding extraction. The final patient-level embedding was obtained by average pooling the embeddings from all chunks.

2.2.4. Radiology Images Processing

The CT scans we had were processed using SMAART-AI, an automated pipeline, to extract the SMA and SM-HU. The SMAART-AI tool computed SM-HU by taking the average Hounsfield value for all pixels marked as skeletal muscle. For the remaining patients, the SMA and SM-HU values extracted by a radiologist using the AW Server [13] were used. SMI was calculated by dividing the SMA (cm$^2$) with the square of height (m$^2$).

For the CT image embeddings, all slices belonging to the third lumbar level (L3) were extracted, using the SMAART-AI tool, for each patient from the venous phase axial series, if available; otherwise, the arterial phase axial series was used. The embeddings of these L3 slices were extracted using RadImageNet [14] through HoneyBee [15]. Average pooling for all the slices per patient was used to get the final embeddings for each patient.

*2.3. Machine Learning Models*

We formulated the problem as a binary classification task, where non-cachectic and pre-cachectic were grouped into non-cachectic, whereas cachectic and refractory were grouped into cachectic.

We employed a Multilayer Perceptron (MLP) model for the classification task, utilizing four hidden layers with dropout regularization. The number of nodes in each layer, dropout values, learning rate, and seed were optimized using Weights & Biases (wandb) [16]. The optimization process involved a Bayesian search strategy, exploring different configurations to determine the optimal set of hyperparameters. Seeds were used to ensure reproducibility of all experiments.

To train and evaluate the model, we implemented 10-fold cross-validation, splitting the training set into 10 subsets, where each subset was used once as a validation set while the remaining data was used for training. The final model performance on the test set was obtained by averaging the predictions across all folds. Five models were trained with different architectures in terms of the number of nodes per layer. The average of the predictions from these five models was used as the final prediction. This approach helped to mitigate overfitting and provided a robust estimate for generalizability. The variance in the predictions from these five models was used to ensure reliability by segregating predictions that were not confident.

*2.4. Experiments*

Model training and testing was carried out on the following combination of data:

1. Clinical data, BMI with SMA, SMI, and SM HU.
2. Clinical data, BMI, SMA, SMI, SM HU, and lab reports.
3. Clinical data, BMI, SMA, SMI, SM HU, lab reports, and structured data are added to the tabular format.

4. Embeddings of the following data modalities were concatenated,
    i. Tabular data: including clinical data, BMI, SMA, SMI, SM HU, and lab reports.
    ii. Clinical notes: only relevant text was used.
5. Embeddings of the following data modalities were concatenated,
    i. Tabular data: including clinical data, BMI, SMA, SMI, SM HU, and lab reports.
    ii. Clinical notes: only relevant text was used.
    iii. Radiology images: CT scan slices of mid-L3.

3. **Results**

*3.1. Comparison of the Model Performance Trained using Various Modalities of Data and Techniques for Combining Data*

The model's prediction accuracy evaluated across different data modalities is presented in Figure 3. The models trained with clinical data and skeletal muscle indices from CT scans have a prediction accuracy of 69%, which improves to 73% with the addition of lab reports and 85% with the addition of structured clinical notes in the training data. The model trained with embeddings of tabular data (including clinical data, SMA, SMI, SM-HU, and lab report) and structured clinical notes have a prediction accuracy of 88%. The confusion matrices and classification reports show the precision and F1 score for cachexia prediction to be 65% and 73% for two data modalities, 69% and 76% with the addition of lab reports, and 80% and 86% with structured notes. Incorporating multimodal embeddings improves cachexia prediction precision and F1 score to up to 92% and 88%.

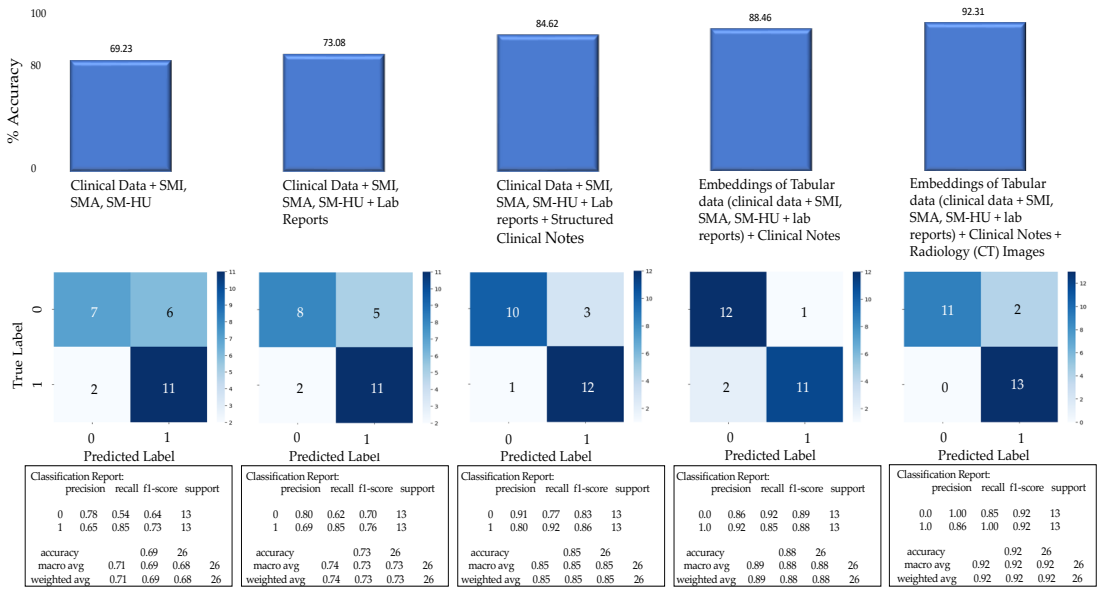

**Figure 3. Model performance comparison using various modalities of data.** Model performance improves with the incorporation of additional data modalities, leading to more informed cachexia predictions. Accuracy, precision, and F1 score further increase when embeddings are included in the training. However, the addition of radiology embeddings appears to reduce accuracy, likely due to the condition of skeletal muscle and adipose tissue in radiology scans supporting the prediction that is not the same as the actual given status.

Figure 4 illustrates the correct and incorrect predictions made by models trained on different combinations of data modalities. The available data modalities for each sample are also provided in the figure. With reference to Figure 4(a), samples 1, 6, and 15 were misclassified by the model trained using clinical data combined with SMA, SMI, and SM HU. The SMI values for these samples aligned with the model's incorrect prediction. However, when lab reports were incorporated into training, the model was able to correctly classify these samples, due to informative lab values that aligned with the correct cachectic status. Samples 8 and 11 were correctly classified by models trained with SMI, but their SMI values were very close to the cutoff. When lab reports were added, the model misclassified these samples, as the lab report values aligned with the incorrect cachexia status. However, clinical notes confirmed the actual cachexia status, leading the model trained with structured clinical notes to make the correct predictions.

In Figure 4(b) samples 13 and 25 were misclassified by models trained with clinical data combined with SM measurements and lab reports, as both modalities supported the incorrect non-cachexia status. However, for sample 13, clinical notes were available and contained evidence supporting cachexia, enabling the model trained with the addition of clinical notes in Figure 4(c) to make the correct prediction. In contrast, sample 25 lacked clinical notes, yet the model trained with clinical notes made the correct prediction, unlike those trained with SM measurements and lab reports.

Sample 22 was cachectic, but SMI and CXI suggested a non-cachectic status. However, SM HU was slightly below the normal range, and neutrophil count was high, both indicating a cachectic tendency. The models trained with SMI and lab reports made correct predictions, but the model trained with clinical notes misclassified the sample, although the clinical notes corresponding to this sample were missing. Both the models trained on embeddings correctly classified the sample.

Sample 26 presented conflicting indicators with SMI slightly below the cutoff and SM HU below the normal range (suggesting cachexia), whereas CXI, neutrophil count, blood urea nitrogen, and creatinine suggested a non-cachectic state. The models trained with SM measurements and lab reports misclassified the sample as cachectic, and the model trained with notes maintained the incorrect classification since notes were missing for this sample. However, both the models trained on embeddings made the correct prediction.

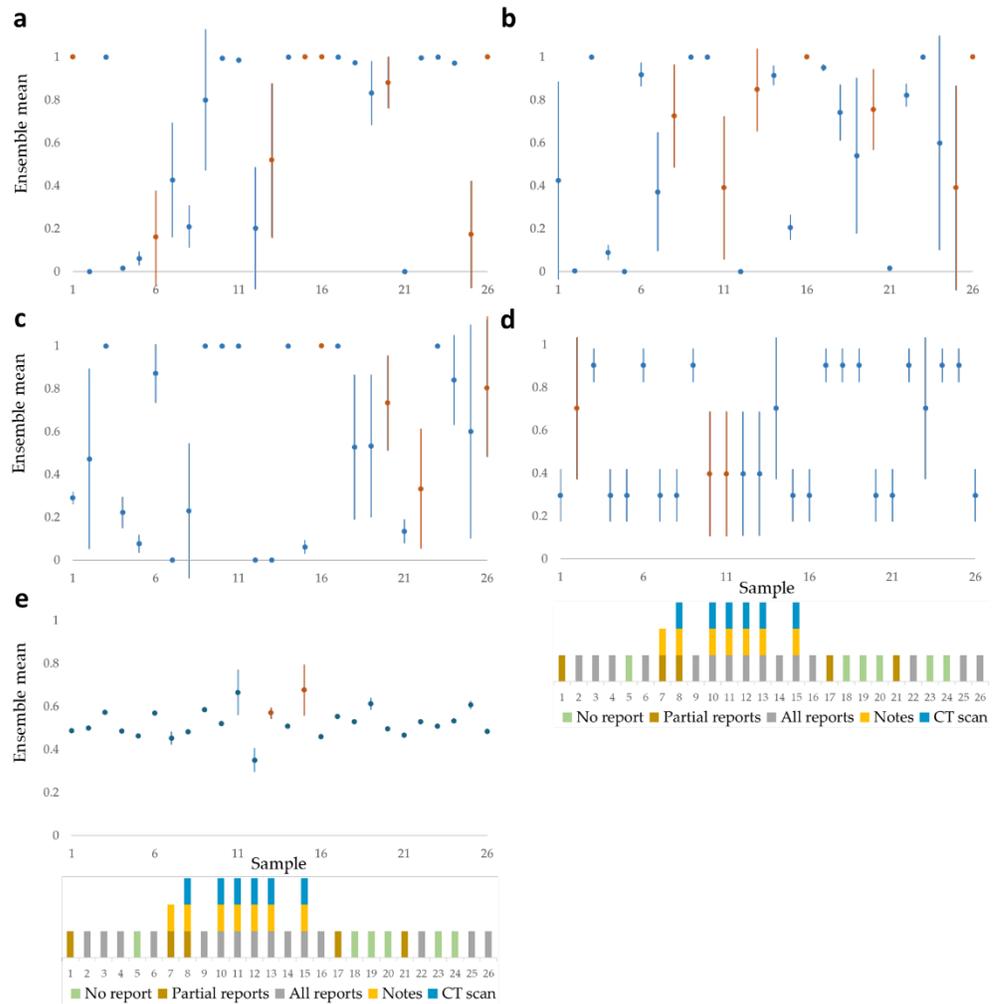

**Figure 4. Comparison of the sample-wise model predictions across different available data modalities**. Additional modalities were combined with clinical data. **(a)** with SM measurements. **(b)** with SM measurements and lab reports. **(c)** with SM measurements lab reports and structured clinical notes in tabular format. **(d)** concatenated embeddings of tabular data (SM measurements and lab reports) and focused clinical notes. **(e)** concatenated embeddings of tabular data (SM measurements and lab reports), focused clinical notes, and CT images corresponding to the third lumbar level. The model is underconfident when the data lacks sufficient evidence across available data modalities to support a confident decision.

*3.2. Model Confidence Analysis*

3.2.1. Model Confidence with Respect to Data Modalities

**Model trained on clinical data with SM measurements:** In Figure 4(a), the model's confidence largely aligns with the available SM measurements (SMI, SM HU) and the actual cachexia status, with a few notable exceptions. Sample 1 was misclassified as non-cachectic with high confidence. This is because SMI was below the cutoff and SM HU was well below the normal range, which strongly supported the incorrect prediction. Samples 2, 3, 4, and 5 were correctly classified, with high confidence, as their SMI values reinforced the model's predictions. Sample 6 was incorrectly classified with low confidence, as its SMI exactly coincided with the cutoff value, making the prediction uncertain. Similarly, samples 7, 8, and 9 were correctly classified, but their SMI values were close to the cutoff, leading to lower model confidence. Sample 12 was predicted as non-cachectic due to SM measurements, CXI, and UCR supporting this status. However, a high neutrophil count (indicative of cachexia) and an unusually high lymphocyte count (not indicative of cachexia) caused the model to be under confident. Samples 15, 16, and 25 had SM measurements that did not indicate cachexia, yet the actual status was cachectic. This led to confident incorrect predictions for samples 15 and 16 and an underconfident incorrect prediction for sample 25. The remaining confident and correct predictions were primarily supported by the SM measurements with a few exceptions.

**Model trained on clinical data with SM measurements and lab reports:** In Figure 4(b), the model's confidence was influenced by conflicting information between SM measurements and lab reports. Sample 1 was correctly classified as cachectic, but with low confidence, due to conflicting SMI and lab report values. Samples 2, 3, and 4 also had conflicting SM and lab report values, but the model maintained high confidence in its correct predictions. Sample 5 had missing lab reports, yet the model remained confident in its prediction based on SM measurements alone. Sample 6 was incorrectly classified, with all values supporting the incorrect decision. However, an unusually high lymphocyte count (unrelated to cachexia) seemed to influence a confident incorrect prediction. Samples 7 and 8 had SM measurements at the cut-off point and sample 7 also had lab reports near the normal range, leading to a correct but low-confidence prediction. Sample 8 had the serum creatinine level slightly below the normal (not indicative of cachexia), but the model used this abnormality to make an incorrect decision with low confidence. Samples 9 and 10 were correctly classified as cachectic, as SMI and high neutrophil counts supported this status. However, CXI was high (suggesting non-cachexia), due to elevated lymphocyte counts. The model likely relied more on neutrophil levels for a confident correct prediction. Samples 11, 13, 20, and 25 had conflicting SM and lab report values, leading to low-confidence predictions. Samples 16 and 26 were incorrectly classified with high confidence because sample 16 had high neutrophil count and low SM measurements that strongly supported the incorrect decision. Sample 26 had low SM measurements, high UCR, and low BUN suggesting muscle loss and malnutrition, leading to a confident incorrect prediction. Sample 19 was correctly classified as non-cachectic, but with low confidence, since SM measurements contradicted the actual status. Sample 24 had SM measurements supporting cachexia, but the model trained with lab reports made a correct prediction with low confidence due to missing lab data.

**Model trained on clinical data with SM measurements, lab reports, and structured clinical notes:** In Figure 4(c), model confidence was influenced by the availability and content of clinical notes. Samples 2, 8, 18, 19, and 25 were correctly classified, but with low confidence, as no clinical notes were available to reinforce the prediction. Sample 8 had notes that partially supported the correct non cachectic status such as unintentional weight loss after surgery, a normal Karnofsky score, and loss of appetite. Samples 1, 3, 4, 5, 6, 9, 14, 17, 21, and 23 had no available notes but maintained confident predictions based on prior knowledge from lab reports, except for sample 1, where the model trained with reports made a low-confidence decision, but the model trained with notes was confident. Samples 7, 10, 11, 12, 13, and 15 had clinical notes supporting the actual status, leading to correct and confident predictions, except sample 13 where the notes partially support cachectic status, but this was not the actual status. Sample 16 remained incorrectly classified, as no clinical notes were available.

**Model trained on embeddings from tabular data (clinical data combined with SM measurements and lab reports) and focused clinical notes:** In Figure 4(d), model confidence was influenced by conflicting information across multiple modalities. Samples 2, 10, 11, 12, 13, 14, and 23 were predicted with low confidence due to conflicting information from different data sources. Samples 14 and 23 were exceptions, where data correctly supported the actual cachexia status, but the model remained underconfident.

**Model trained on embeddings from tabular data (clinical data combined with SM measurements and lab reports), focused clinical notes, and radiology images (CT scans):** In Figure 4(e) model decisions are influenced by information from CT images.

This analysis considers the additional data modalities (SM measurements, lab reports, and clinical notes) rather than the clinical data, which also impacted the model's overall confidence in its decision.

3.2.2. Model Confidence Correct versus Incorrect Predictions

Figure 5 presents the quantified confidence of correct versus incorrect predictions compared to models trained on various data modalities. In all models, incorrect predictions exhibit higher mean and median variance, indicating lower confidence, compared to correct predictions. As additional data modalities are incorporated, the mean and median variance of correct predictions gradually increase, with the model trained on embeddings showing the highest values. Notably, the variance distributions of correct and incorrect predictions are well separated in the embedding-based model. This clear distinction in mean and median variance values suggests that incorrect predictions can be effectively identified using a threshold on quantified variance.

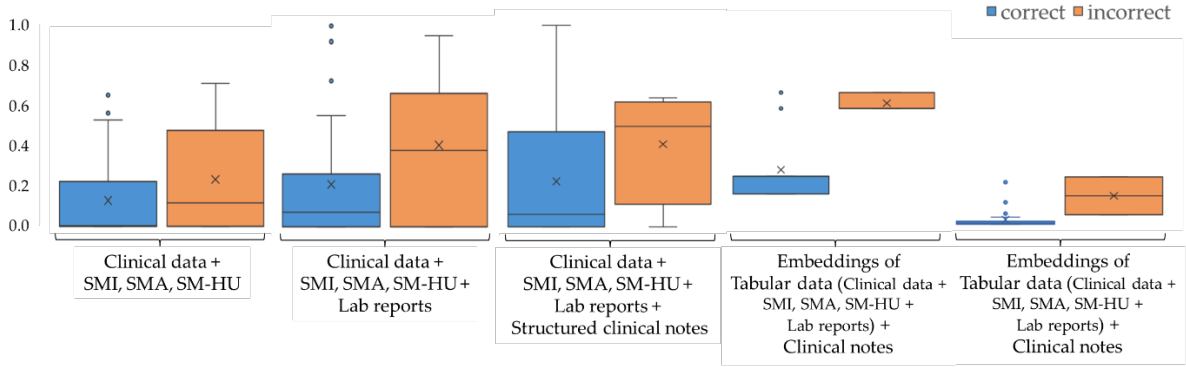

**Figure 5. Comparison of confidence levels in correct vs. incorrect decisions across models trained on different data modalities.** Correct predictions have a lower mean and median variance, which means that the model is generally more confident when making correct decisions compared to incorrect ones. Although there are certain correct predictions on which the model is not confident.

*3.3. Performance Comparison of Large Language Models for Extracting Structured Data from Clinical Notes*

Overall Deepseek performed better with an average score of 24.6 (94.62%) followed closely by Qwen with an average score of 23 (88.46%). Llama3.2 has the lowest score of 21.2 (81.54%). A few samples have close scores from all three models.

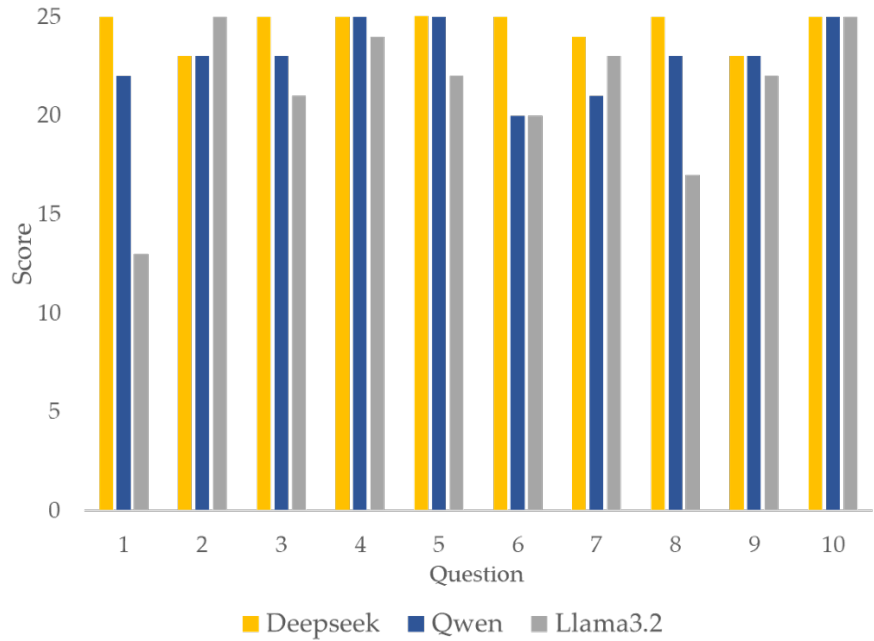

**Figure 6. Performance comparison of the different LLMs used for extracting structured data from clinical notes.** Out of the clinical notes of 105 patients, a random sample of 10 patients was selected, and the performance was scored by comparing the response in 'yes', 'no', and 'not given' of each model for the set of 26 questions. The Deepseek model performed the best for this task, followed closely by Qwen.

4. **Discussion**

Cancer cachexia is a complex, multifactorial syndrome that lacks a single definitive biomarker, making its identification inherently multimodal. Various indicators, ranging from skeletal muscle loss in radiological scans to metabolic disruptions in lab reports and clinical notes, contribute to its diagnosis. Additionally, patient demographics and overall medical condition further influence its manifestation. Given this complexity, integrating all available patient data is crucial for early and accurate detection, ensuring timely intervention and better patient outcomes.

A significant amount of valuable information is contained within clinical notes, which often describe the physical assessment of muscle loss, performance status, previous weight history, unintentional weight loss, and anemia. These notes may also capture subjective patient experiences, such as feelings of satiety or fullness, nausea, anorexia, diarrhea, and psychiatric conditions, all of which are highly relevant for accurately detecting cancer cachexia. Traditionally, extracting insights from clinical notes has been challenging due to their unstructured nature, but the recent advancement in large language models (LLMs) now enable the extraction of relevant information in its original form or as structured data, making it possible to integrate these insights into predictive models.

Foundation models trained on radiological images and EHR data generate enriched embeddings that have been used to learn downstream tasks, such as cachexia detection. Embedding-based learning provides a powerful framework to extract and integrate diverse data sources [17-19]. Leveraging multimodal models have been shown to improve cachexia detection efficiency.

The findings of this study emphasize that incorporating structured, unstructured, and imaging data, including electronic health records (EHRs), lab reports, clinical notes, and radiology images, makes the ML model more aware to a patient's overall health status in relation to cachexia, leading to improved predictive accuracy. When multiple data sources align, the model gains higher confidence and reliability in its predictions [17, 20]. However, when data sources conflict, the model's confidence decreases, signaling cases that require closer clinical evaluation. Model confidence is therefore crucial not only for detecting potential misclassifications but also for identifying cases that may benefit from expert review. Even when the model makes a correct prediction with low confidence, examining conflicting indicators can offer valuable insights into the complexities of cancer cachexia and its interactions with other conditions.

In real-world clinical settings, comprehensive patient data is often incomplete, posing a challenge for ML-based approaches[21]. Therefore, for an AI-driven solution to be clinically viable, it must be able to handle missing data, ensuring reliable predictions even when certain modalities are unavailable. By building models that can adapt to real-world constraints, this research moves closer to a practical and scalable solution for detecting and managing cancer cachexia in clinical practice.

5. Conclusion

This study leverages open-source LLMs and foundation models trained on medical data to integrate diverse patient information, enabling the development of multimodal models for the early detection of cancer cachexia. Additionally, our prediction confidence estimation helps identify cases that require expert analysis, making the solution reliable. A multimodal ML framework that effectively utilizes available clinical data holds significant potential as a real-world, scalable solution, assisting clinicians in diagnosing, monitoring, and managing cancer cachexia throughout a patient's treatment journey.

**Author Contributions:** Conceptualization, S.A. and G.R.; methodology, S.A. and G.R.; software, S.A.; validation, S.A., N.P., M.P., E.D., J.P., M.S., Y.Y. and G.R.; resources, N.P., M.P., J.P., and G.R.; data curation, S.A.; writing—original draft preparation, S.A.; writing—review and editing, N.P., M.P., E.D., J.P., M.S., Y.Y. and G.R.; visualization, S.A.; supervision, G.R. and Y.Y.; project administration, G.R.; funding acquisition, G.R., J.P, M.S. All authors have read and agreed to the published version of the manuscript.

**Funding:** This research was funded by NSF grants 2234468 and 2234836, NIH grant U01CA200464, James and Esther King Foundation grant 8JK02, and Department of Defense grant PA210192.

**Institutional Review Board Statement:** The study was conducted in accordance with the Declaration of Helsinki and approved by the Institutional Review Board (or Ethics Committee) of Moffitt Cancer Center MCC 22299, MCC 19717 (version 1.4 approved on 02/12/2019), and MCC21962, MCC20105.

**Informed Consent Statement:** Informed consent was obtained from all subjects involved in the study.
**Conflicts of Interest:** The authors declare no conflicts of interest.

1. Han, J., et al., *Imaging modalities for diagnosis and monitoring of cancer cachexia.* EJNMMI research, 2021. **11**: p. 1--18.
2. Araújo, J.P., et al., *Nutritional markers and prognosis in cardiac cachexia.* International journal of cardiology, 2011. **146**(3): p. 359-363.
3. Fearon, K., et al., *Definition and classification of cancer cachexia: an international consensus.* The lancet oncology, 2011. **12**(5): p. 489-495.
4. Go, S.I., et al., *Cachexia index as a potential biomarker for cancer cachexia and a prognostic indicator in diffuse large B-cell lymphoma.* Journal of cachexia, sarcopenia and muscle, 2021. **12**(6): p. 2211-2219.
5. Yuan, Q., et al., *Developing and validating a modified cachexia index to predict the outcomes for colorectal cancer after radical surgery.* European Journal of Clinical Nutrition, 2024. **78**(10): p. 880-886.
6. Argilés, J.M., et al., *The cachexia score (CASCO): a new tool for staging cachectic cancer patients.* Journal of cachexia, sarcopenia and muscle, 2011. **2**: p. 87-93.
7. Chen, Y., et al., *Machine learning to identify precachexia and cachexia: a multicenter, retrospective cohort study.* Supportive Care in Cancer, 2024. **32**(10): p. 630.
8. Permuth, J.B., et al., *The Florida pancreas collaborative next-generation biobank: infrastructure to reduce disparities and improve survival for a diverse cohort of patients with pancreatic cancer.* Cancers, 2021. **13**(4): p. 809.

9. Fearon, K., et al., *Definition and classification of cancer cachexia: an international consensus.* The lancet oncology, 2011. **12**(5): p. 489--495.
10. Vigano, A.A.L., et al., *Use of routinely available clinical, nutritional, and functional criteria to classify cachexia in advanced cancer patients.* Clinical nutrition, 2017. **36**(5): p. 1378--1390.
11. Yang, X., et al., *A large language model for electronic health records.* NPJ digital medicine, 2022. **5**(1): p. 194.
12. Jafri, S.H.R., et al., *Cachexia index in advanced non-small-cell lung cancer patients.* Clinical Medicine Insights: Oncology, 2015. **9**: p. CMO. S30891.
13. Beenish Zia, S.A., Steve Johnson, Jared Pager, S Antoine Aliotti, Yingpo Huang, Jerome Knoplioch, Yannick Leberre, Lionel Marmonier, Florentin Toulemon, *White Paper: Accelerate Your Visualization Experience.* . 2020, Intel Corporation and GE Healthcare: Santa Clara, CA.
14. Mei, X., et al., *RadImageNet: an open radiologic deep learning research dataset for effective transfer learning.* Radiology: Artificial Intelligence, 2022. **4**(5): p. e210315.
15. Tripathi, A., et al., *Honeybee: a scalable modular framework for creating multimodal oncology datasets with foundational embedding models.* arXiv preprint arXiv:2405.07460, 2024.
16. Biewald, L., *Experiment Tracking with Weights and Biases.* 2020.
17. Waqas, A., et al., *Embedding-based Multimodal Learning on Pan-Squamous Cell Carcinomas for Improved Survival Outcomes.* arXiv preprint arXiv:2406.08521, 2024.
18. Waqas, A., et al., *Multimodal data integration for oncology in the era of deep neural networks: a review.* Frontiers in Artificial Intelligence, 2024. **7**: p. 1408843.
19. Waqas, A., et al., *Revolutionizing digital pathology with the power of generative artificial intelligence and foundation models.* Laboratory Investigation, 2023: p. 100255.
20. Waqas, A., et al., *SeNMo: A self-normalizing deep learning model for enhanced multi-omics data analysis in oncology.* arXiv preprint arXiv:2405.08226, 2024.
21. Tripathi, A., et al., *Building flexible, scalable, and machine learning-ready multimodal oncology datasets.* Sensors, 2024. **24**(5): p. 1634.